\documentclass[prl,aps,twocolumn,showpacs]{revtex4}
\usepackage{graphicx}
\usepackage{dcolumn}
\usepackage{amsmath}

\newcommand{\kB}{k_{\mathrm{B}}}
\newcommand{\MB}[1]{{\mbox{\mathversion{bold}$#1$}}}
\newcommand{\EXP}[1]{\,e\mbox{\raisebox{1,3ex}{$\,#1$}}\,} 

\begin{document}

\title{Transition to linear domain walls in nano-constrictions}

\author{N.\ Kazantseva, R.\ Wieser, U.\ Nowak}

\affiliation{Fachbereich Physik, Universit\"at Duisburg-Essen,
  47048 Duisburg, Germany}

\begin{abstract}
Domain walls in nano-constrictions are investigated with a focus on
thermal properties. In general, the magnetization component
perpendicular to the easy axis which in a domain wall usually occurs
has a value different from the easy-axis bulk magnetization value with
a separate phase transition at a critical temperature below the Curie
temperature. Since this effect is the more pronounced the smaller the
domain wall width is we investigate it especially in domain walls with
a confined geometry, using analytical arguments, mean-field theory and
Monte Carlo simulations. Our findings may contribute to the
understanding of magneto-resistive effects in domain walls with sizes
of only a few atomic layers, as e.\ g.\ in nano-contacts or
nano-constrictions.
\end{abstract}


\maketitle

Well controlled domain walls could become important constituents of
future magneto-electronic devices \cite{allwoodSCIENCE02}. Especially,
the understanding of domain walls in confined nanometric geometries is
important since those can show behavior deviating from their usual
bulk properties, as e. g. a controlled pinning and a strongly reduced
domain wall width
\cite{brunoPRL99,pietzschPRL00,miyakeJAP02,klauiPRL03}. The latter is
thought to contribute to large magneto-resistance effects of domain
walls in nano-wires \cite{ebelsPRL00}, nano-constrictions
\cite{ruesterPRL03}, and nano-contacts \cite{garciaPRL99,chopraPRB02}.

In only few publications the temperature dependence of domain wall
properties was investigated
\cite{garaninPA91b,labayeJAP02,coeyPRB03}. Most important in this
context is the pioneering work of Bulaevski and Ginzburg
\cite{bulaevskiiJETP64} who showed within the framework of
Ginzburg-Landau-theory that for a one dimensional domain wall profile
(e.\ g.\ a Bloch wall) the easy-axis and hard-axis components of the
magnetization, respectively, are two separate order parameters with
different critical temperatures. In other words, the perpendicular
magnetization component which arises necessarily in a domain wall has
at finite temperatures values lower than the easy-axis equilibrium
magnetization (leading to the term ``elliptical domain walls'') and
vanishes completely for a temperature $T_h$ which is lower than the
Curie temperature $T_c$ of the bulk material (leading to the term
``linear domain walls'' for temperatures $T_h < T < T_c$ ).

However, the deviation of $T_h$ from $T_c$ is proportional to the
squared inverse domain wall width \cite{bulaevskiiJETP64} and, hence,
should be very small.  Consequently, linear walls are hard to detect
experimentally \cite{koetzlerPRL93}, their relevance is considered to
be rather low and most of the numerical calculations of domain wall
properties are performed using micromagnetic codes where the
assumption of a constant magnetization value is made in contradiction
to the findings described above. In the following we will investigate
in how far thermodynamic deviations from pure Bloch-like domain wall
structures can become relevant due to the reduced size of domain walls
in confined geometries.  This is an important question since it was
suggested \cite{dzeroPRB03} that linear domain walls might explain the
observed large magneto-resistance effects in constrained domain walls.

\begin{figure}
  \begin{center} 
    \includegraphics[width=7.5cm]{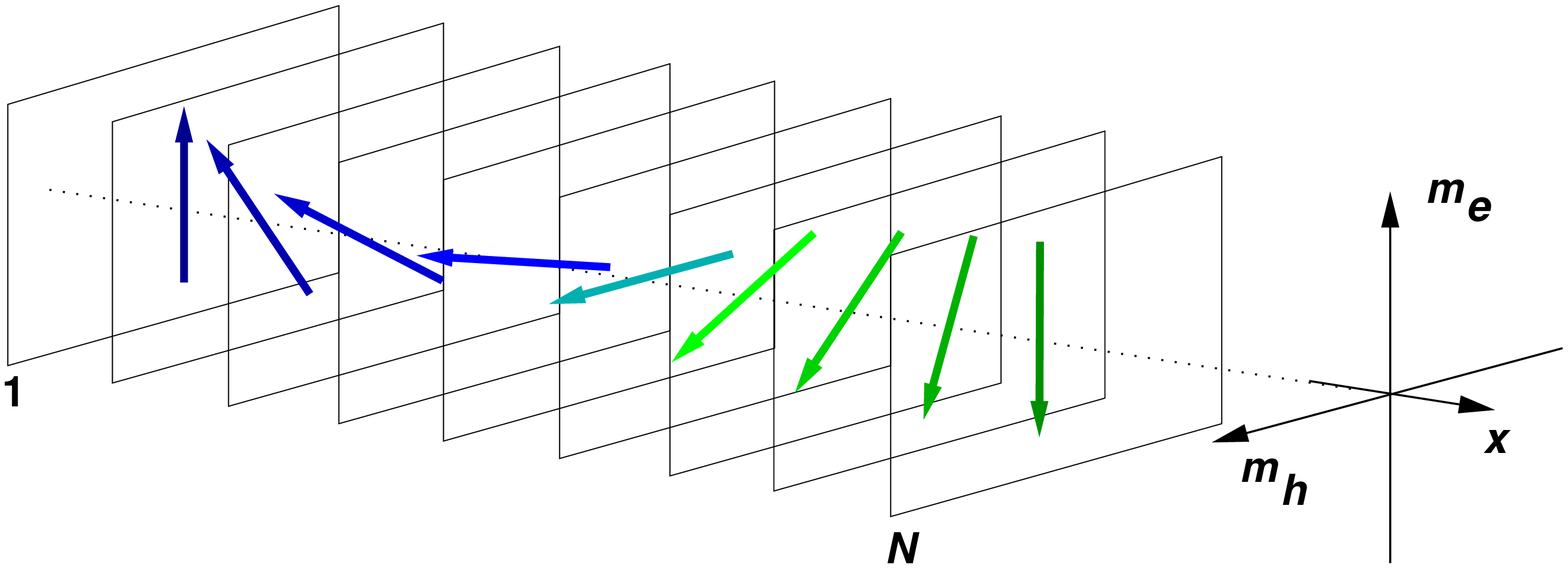}
  \end{center} 
  \caption{Sketch of the wall geometry. The magnetization of the first
  and the last plane are antiparallel along the easy axis. The
  magnetization rotates in a plane via the hard axis.}
  \label{f:scetch}
\end{figure}

In the following we consider a domain wall structure as shown in
Fig. \ref{f:scetch}. In the picture, each arrow represents the mean
magnetization of a plane. Our Monte Carlo (MC) simulations are for a
full three dimensional model allowing for magnetization fluctuations
within the planes while in the mean-field (MF) approximation each
(infinite) plane is represented as one magnetization vector so that
the model is effectively one dimensional. Depending on the details of
the methods we apply we use either fixed boundary conditions where the
first and the last plane of the system are fixed as shown in the
figure or we use anti-periodic boundary conditions. In both cases we
force a domain wall into the system which for large system size
(number of planes $N$) will not fill up the whole system. However, for
smaller system size the boundary conditions force the domain wall to
adopt the system size. This models a domain wall caught in a
nano-constriction \cite{brunoPRL99} where perfect pinning is assumed.

We investigate the system in terms of a classical spin model with
spin variables $|\MB{S}_i| = 1$ on a cubic lattice with lattice
constant $a$ and energy contributions from ferromagnetic exchange
between nearest neighbors with coupling constant $J$ and a uniaxial
anisotropy with $D > 0$ defining the easy axis of the system,
\begin{eqnarray}
  {\cal H} = & - & \frac{J}{2}\sum_{\langle i,j
  \rangle}\MB{S}_i\cdot\MB{S}_j - D\sum_i\left(S_i^e\right)^2. 
  \label{e:ham}
\end{eqnarray}

In the zero temperature limit all spins within a plane will be
parallel and the domain wall profiles can be calculated analytically
in the continuum limit where the energy density (per cross-sectional
area) is a one dimensional integral
\[
  e \; = \; \frac{J}{2 \, a} \; \int\limits_{-L/2}^{L/2} \left (
  \nabla\cdot{\bf S} \right )^2 {\rm d} x\;-\;
  \frac{D}{a^3}\;\int\limits_{-L/2}^{L/2} \left( S^e\right)^2 {\rm
  d}x.
\]
The magnetization of the domain wall rotates within a plane and can be
expressed by a single angle of rotation $\theta$. The
Euler-Lagrange-equation which minimizes the energy above is solved by
an elliptic integral
\[
  x\,= \,\int\limits_{-\pi/2}^\phi\frac{\delta_0 {\rm
  d}\theta}{\sqrt{c^2-\sin^2\theta}} = \,\frac{\delta_0}{c}{\rm
  F}\left(\phi, \frac{1}{c^2}\right)
\]
where $\delta_0 = a \sqrt{J/2D}$ is the zero-temperature domain wall
width of an unconstrained wall and the integration constant $c$ is
given by the boundary condition $x(\pi/2) = L/2$ with $L = (N-1) a$.
The wall profiles are then given by Jacobian sine and cosine functions
\begin{equation}
  S_e(x) \, = \,{\rm sn} \Big(\frac{c x}{\delta_0},
  \frac{1}{c^2}\Big) \quad\quad S_h(x) \, = \,{\rm cn}\Big(\frac{c
  x}{\delta_0}, \frac{1}{c^2}\Big).
  \label{e:jacob}
\end{equation}
The limit $c \to 1$ corresponds to an unconstrained wall with the
usual Bloch-wall profiles
\begin{equation}
  S_e(x) \, = \, \tanh (x / \delta_0)  \quad\quad S_h(x)
  \, = \, \cosh^{-1} (x / \delta_0 ).
  \label{e:hyper}
\end{equation}
The opposite case, $c \gg 1$, corresponds to a very constrained wall
($L \ll \delta_0$) where the wall is forced to adapt the system size
and the profiles follow simple trigonometric functions,
\begin{equation}
  S_e(x) \, = \, \sin (\pi x/L ) \quad\quad S_h(x) \, = \, \cos (\pi x/L).
  \label{e:trig}
\end{equation}
In this limit the actual domain wall width is $L$.

\begin{figure}[ht]
  \begin{center}
    \includegraphics[angle = 270, width=7cm]{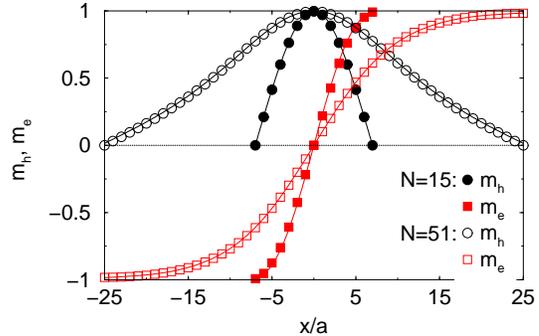}
  \end{center}
  \caption{Ground state easy- and hard-axis magnetization profiles for
  constrained walls. Data points are from low temperature MF
  calculations, solid lines correspond to
  Eq. \ref{e:jacob}. $D/J=0.003$.}
  \label{f:constraint}
\end{figure}
As an example, Fig. \ref{f:constraint} shows a comparison of these
expressions numerical MF results, obtained as described below in the
low temperature limit $\kB T = 0.02J$. The smaller system shows the
case of a very constrained wall where the wall profiles are already
described by trigonometric functions while the larger system shows an
intermediate case.  Note that in Ref. \cite{brunoPRL99} corresponding
calculations were made, but for other boundary conditions and system
geometries, respectively.

However, we want to focus on thermal properties and in order to obtain
results for finite temperatures we start with the MF Hamilton
operator which (neglecting terms without $\MB{S}$) is
\begin{equation}
  {\cal H}_{\rm MF} = - \sum_{i} \Big( J \MB{S}_i \cdot (\MB{m}_{i-1} \! + \!
  4 \MB{m}_i \! + \! \MB{m}_{i+1} ) + D \left(S_i^e\right)^2 \Big)
  \label{e:mf-ham}
\end{equation}
where the contribution $4\MB{S}_i \cdot \MB{m}_i$ comes from the four
neighbors within the plane. Then we solve the MF
self-consistency equations,
\begin{eqnarray}
  {\mathbf m}_i = \langle {\mathbf S}_i \rangle = \frac{1}{Z} \;
  \mathrm{Tr} \; {\mathbf S}_i\EXP{-{\cal H}_{\rm MF}/\kB T} \nonumber
\end{eqnarray}
with $Z = \mathrm{Tr} \EXP{-{\cal H}_{\rm MF}/\kB T}$
numerically. Here, ${\mathbf m}_i$ is the thermally averaged
magnetization of the $i$th plane and the trace is an integral over the
unit sphere. These equations can be solved iteratively, starting with
an arbitrary magnetization profile and then let the equations evolve
until a stationary state is reached.

\begin{figure}
  \begin{center}
    \includegraphics[angle=270, width=7cm]{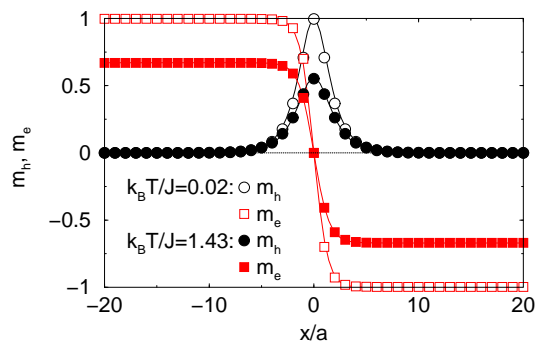}
  \end{center}
  \caption {Easy- and hard-axis magnetization profiles in MF
    approximation for two different temperatures. $D/J = 0.03$}
  \label{f:profile}
\end{figure}

Let us start with the case of an unconstrained wall.
Fig. \ref{f:profile} shows domain wall profiles for two different
temperatures where the system size $L = 40a$ is large enough so that
an equilibrium domain wall for an anisotropy value of $D = 0.03J$ fits
well into the system.  The solid lines are the analytical functions as
calculated above, here, simply the tanh and $1/ \cosh$
profiles. Interestingly, for finite temperatures the mathematical form
of the wall profile is conserved, solely the amplitudes and the domain
wall widths vary with temperature. Hence, in the limit $L \gg
\delta_0$ the thermodynamic wall profiles can be described as in
Eqs. \ref{e:hyper} but with a temperature dependent domain wall width
$\delta(T)$ and temperature dependent amplitudes $M_h(T)$ and
$M_e(T)$. Furthermore, Fig. \ref{f:profile} suggests that that the
temperature dependence of the two amplitudes is not the same.

\begin{figure}
  \begin{center}
  \includegraphics[angle=270, width=7cm]{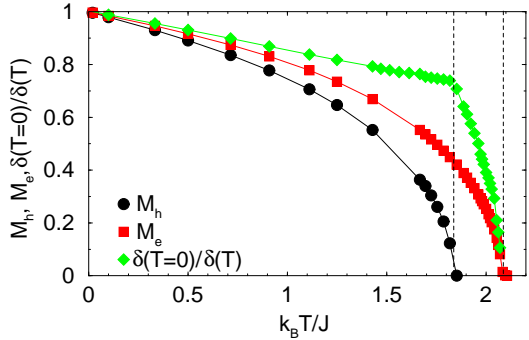}
  \end{center}
  \caption{Temperature dependence of easy- and hard-axis order
    parameters and (reduced) inverse domain wall width from MF
    calculations. $L = 50a$, $D/J= 0.3$.}
  \label{f:order}
\end{figure}

Instead, these amplitudes, $M_h$ and $M_e$, define two distinct order
parameters as shown in Fig. \ref{f:order}. Obviously, these two order
parameters vanish continously at two different temperatures where the
upper one is the usual Curie temperature $T_c$ and the lower one is a
second critical temperature, $T_h$, which describes the phase
transition of the hard-axis component of the magnetization vector. For
temperatures $T_h < T < T_c$ the domain wall is linear with an
easy-axis component of the magnetization only. For $T<T_h$ in general
one finds an elliptical wall profile which for lower temperature goes
over to the usual Bloch wall profile. Also shown in Fig. \ref{f:order}
is the reduced, inverse domain wall width demonstrating that
$\delta(T)$ increases slightly with temperature for elliptical domain
walls, shows a kink at $T_h$, and then diverges in the linear wall
regime approaching $T_c$.

\begin{figure}
  \begin{center} 
    \includegraphics[angle=270, width=7cm]{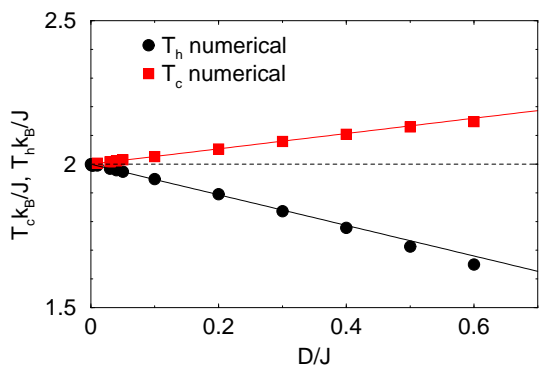}
  \end{center} 
\caption{Anisotropy dependence of the two critical temperatures $T_c$
  and $T_h$. Data points are from MF calculations as in
  Fig. \ref{f:order}, solid lines from Eq. \ref{e:dtc}.}
  \label{f:tcs}
\end{figure}
Both critical temperatures depend on the strength of the anisotropy
$D$ as is shown in Fig. \ref{f:tcs}. $T_c$ corresponds to the usual
bulk Curie temperature and its anisotropy dependence expanded with
respect to $D/J$ is $\kB T_c / J = 2 + \frac{4}{15} D/J + {\cal O}
[D/J]^2$. As Fig. \ref{f:tcs} demonstrates, $T_h$ has also a linear
dependence on $D/J$ in the range of anisotropies which is shown in the
figure and our numerical results suggest $\kB T_h/J \approx 2 - 0.53
D/J$.  This means that the difference between both critical
temperatures scales with the squared inverse domain wall width
\begin{equation}
  \kB (T_c -T_h) /J \approx 0.40 a^2 / \delta_0^2.
  \label{e:dtc}
\end{equation}

These findings, the second phase transition, its dependence on
$\delta_0$ and the diverging domain wall width are qualitatively in
agreement with the earlier calculation within the framework of the
Ginzburg-Landau theory \cite{bulaevskiiJETP64} --- an expansion close
to $T_c$ --- where it was also shown that close to $T_c$ the (linear)
wall has a tanh-profile.  However, for experimental systems reasonable
anisotropies are rather small, so that in general the two critical
temperatures should nearly coincide and $M_h$ should be close to the
easy-axis magnetization $M_e$. Nevertheless, experimental
investigations of linear domain walls exist. In \cite{koetzlerPRL93}
the influence of the wall structure (either elliptical or linear) on
the domain wall mobility was investigated. Here, the deviation of
$T_h$ from $T_c$ was 1\%.  However, since --- as Eq. \ref{e:dtc} shows
--- these effects increase with decreasing domain wall width larger
effects should occur in smaller domain walls as found, e.\ g.\, in
constrained geometries.

For a strongly constrained domain wall and in MF approximation $T_h$
can be estimated analytically. For a system with antiperiodic boundary
conditions and small anisotropy, so that $L \ll \delta_0$, the wall
profiles follow trigonometric functions which means that the angle of
rotation from plane to plane changes in each plane by the same amount
$\pi a /L$. This leads to a decrease of the MF coming from the two
adjacent planes proportional to $\cos(\pi a /L)$. Including this in
the MF Hamiltonian (Eq. \ref{e:mf-ham}) the critical temperature for
zero anisotropy can be calculated in the usual way, now leading to
\begin{equation}
  \kB T_h / J \, = \, \frac{4}{3} + \frac{2}{3} \cos(\frac{\pi a}{L}) .
  \label{e:dth}
\end{equation}
Note, that for a constrained domain wall with $L \ll \delta_0$ much
bigger effects may occur than before (Eq. \ref{e:dtc}).
\begin{figure}[ht]
  \begin{center}
  \includegraphics[angle = 270,width=7cm]{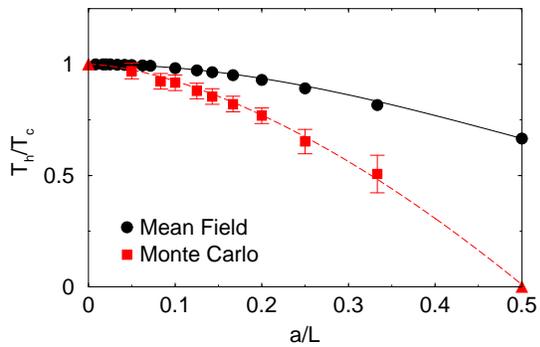}
  \end{center}
  \caption{Dependence of the critical temperature $T_h$ from the size
  of the constrained domain wall. Comparison of MC data, MF
  calculations and Eq. \ref{e:dth} (solid line). The dashed line is a
  guide to the eye.}
  \label{f:tcs2}
\end{figure}

In Fig. \ref{f:tcs2} we compare the formula above with numerical
calculations for the case of constrained walls. We use $D=0$, so that
the condition $L \ll \delta_0$ is always fulfilled, and simulate
systems with fixed boundary conditions calculating $T_h$ from the
hard-axis magnetization. Fig. \ref{f:tcs2} demonstrates that
pronounced effects can be found when the domain wall is constrained to
only a few atomic layers. The agreement of our MF data with
Eq. \ref{e:dth} is very good.  Nevertheless, the MF approximation
strongly underestimates this size dependent effect: this can be
concluded from the fact that even in the extreme case of a three layer
system --- which due to the fixed boundary condition corresponds to a
free monolayer --- a finite $T_h$ is found even though it is clear
from the Mermin-Wagner theorem that no order should occur.

Therefore, we additionally used MC methods to investigate the
break-down of ferromagnetic order in the original spin-model
(Eq. \ref{e:ham}) more rigorously. Using MC methods thermal
excitations (thermally excited spin-waves) are fully taken into
account.  We use a heat-bath algorithm and single-spin-flip dynamics
for the simulations. At every MC step each spin is subject to a trial
step consisting of a small deviation from the original direction
\cite{nowakARCP01}. The lateral dimension of the system is up to $128
\times 128$ with periodic boundary conditions. The number of planes is
varied from 3 to 21 where we use fixed boundary conditions for the
first and last planes.  We start the simulations with an abrupt domain
wall and let it relax for 10000MCS.  Then we calculate the absolute
value of the magnetization component perpendicular to the easy axis,
averaged over the whole system and for another 100000MCS as order
parameter of the phase transition. Note that the precise definition of
the order parameter is important: without calculating the absolute
value of the perpendicular magnetization the long time average of one
magnetization component will always be zero since the wall
magnetization might rotate in the hard plane. Furthermore, to average
over the whole system is also important since the wall can move
diffusively.

$T_h$ is then determined from a finite-size scaling analysis of the
order parameter above where the lateral system size is varied from $8
\times 8$ to $128 \times 128$. The resulting data points are also
shown in Fig. \ref{f:tcs2}. The scaling analysis works well with the
exponents $ \beta$ and $\nu$ from the three dimensional Heisenberg
model. Only for a very small number of planes deviations occur
indicating a crossover from three to two dimensional behavior
\cite{garaninJPA96}.  As expected $T_h$ now goes to zero in the
limiting case of a tri-layer system so that the change of the critical
temperature is more dramatic. E. g., for a wall consisting of 5 atomic
planes ($L=4a$) $T_h$ is reduced by about 35\% as compared to the
Curie temperature and even for a temperature of 0.5$T_c$ the hard-axis
magnetization (the degree of order within the wall) is reduced by
about 20\% as compared to the easy-axis bulk value.

To summarize, investigating the influence of thermal activation on the
properties of domain walls in nano-constrictions we have demonstrated
that with increasing temperature Bloch wall profiles change via
elliptical walls to linear domain walls. The temperature range where
these effects occur scales with the squared inverse domain wall width
so that in general it is rather small. However, since in confined
geometries the relevant quantity is the size of the constriction
larger effects can be found. The break-down of ferromagnetic order is
due to the fact that in a domain wall the mean exchange field
decreases due to the finite angle of rotation between neighboring
magnetic moments. Hence, it is a general effect which will also occur
in other types of domain walls as, e. g., vortex walls. Our findings
may have an impact on the understanding of domain wall
magneto-resistance properties
\cite{dzeroPRB03,levyPRL97,yavorskyPRB02,bergeretPRB03}, especially its
temperature dependence, for two reasons: first, in an elliptical or
linear domain wall the degree of spin disorder is larger in the wall
than in the bulk of the domain since the value of the order parameter
is lower. Second, as suggested in \cite{dzeroPRB03}, in a linear
domain wall the change of the magnetization direction is abrupt while
only its value is changing. This means that the spin of a conductance
electron passing a linear wall cannot follow a continuously rotating
magnetization direction.

\acknowledgments The authors thank R. W. Chantrell and K. D. Usadel
for stimulating discussions. Work supported by the {\em Deutsche
Forschungsgemeinschaft} (SFB 491).


\end{document}